\documentclass[aps,pra,twocolumn]{revtex4-2}
\usepackage{todonotes}
\usepackage{soul}
\usepackage{amsmath,epsfig,mathrsfs,graphicx,amssymb,color}
\usepackage[T1]{fontenc}
\usepackage{t1enc}
\usepackage{hyperref}
\usepackage{array}
\usepackage{booktabs}

\def\be#1\ee{\begin{equation}#1\end{equation}}
\newcommand{\ba}{\begin{eqnarray} }
\newcommand{\ea}{\end{eqnarray} }
\def\cor#1{{#1}}

\usepackage[cp1250]{inputenc}   
\usepackage{amsmath}            
\usepackage{amsfonts}           
\usepackage{amssymb}            
\usepackage{enumerate,graphicx}
\usepackage{array}
\usepackage{booktabs}

\usepackage{physics}

\usepackage{amsmath}            
\usepackage{amsfonts}           
\usepackage{amssymb}            
\usepackage{enumerate,graphicx}
\usepackage{array}
\usepackage{booktabs}
\usepackage{yquant}
\usetikzlibrary{fit, quotes}
\usetikzlibrary{snakes,arrows,shapes}
\useyquantlanguage{groups}
\usepackage{physics}

\def\mb{\begin{pmatrix}}
\def\me{\end{pmatrix}}
\def\be#1\ee{\begin{equation}#1\end{equation}}

\def\mb{\begin{pmatrix}}
\def\me{\end{pmatrix}}
\def\be#1\ee{\begin{equation}#1\end{equation}}

\begin{document}


\title{\cor{Device-independent prepare-and-prepare bipartite null witness dimension test with a single joint measurement}}

\author{Josep Batle$^{1,2}$}
\email{jbv276@uib.es, batlequantum@gmail.com}         
\author{Tomasz Bia{\l}ecki$^{3,4}$}
\author{Tomasz Rybotycki$^{5,6,7}$}
\author{Adam Bednorz$^{3}$}
\email{Adam.Bednorz@fuw.edu.pl}

\affiliation{$^1$
Departament de F\'isica and Institut d'Aplicacions Computacionals de Codi Comunitari (IAC3),
Campus UIB, E-07122 Palma de Mallorca, Balearic Islands, Spain}

\affiliation{$^2$CRISP -- Centre de Recerca Independent de sa Pobla, 07420, sa Pobla, Balearic Islands, Spain}
\affiliation{$^3$Faculty of Physics, University of Warsaw, ul. Pasteura 5, PL02-093 Warsaw, Poland}
\affiliation{$^4$University of Lodz, Faculty of Physics and Applied Informatics,
ul. Pomorska 149/13, PL90-236 {\L}odz, Poland} 
\affiliation{$^5$Systems Research Institute, Polish Academy of Sciences, ul. Newelska 6, 01-447 Warsaw, Poland}
\affiliation{$^6$Nicolaus Copernicus Astronomical Center, Polish Academy of Sciences, ul. Bartycka 18, 00-716 Warsaw, Poland }
\affiliation{$^7$Center of Excellence in Artificial Intelligence, AGH University, al. Mickiewicza 30, 30-059 Cracow, Poland }

\begin{abstract}
We propose a device-independent \cor{null witness} dimensionality test with bipartite measurements and input from two separate parties.
The dimension is determined from the rank of the matrix of measurements for pairs of states prepared by the parties. 
We have applied the test to various IBM Quantum devices. The results demonstrate extreme precision of the test, which is able to detect disagreements with the qubit (two-level) space of bipartite measurement even in the presence of technical imperfections.
The deviations beyond 6 standard deviations have no simple origin and need urgent explanations to unblock progress in quantum computing.
\end{abstract}

\maketitle

\section{Introduction}
Quantum bipartite systems have nonclassical properties, they violate local realism \cite{epr,bell,chsh,chi,eber} 
and are useful in quantum computation. The very, if not the most, important part of quantum operations is measurement.
The measurement often completes the dynamics and reflects the knowledge about the quantum state. A single measurement does not reveal
the full quantum state and gives a random outcome, is invasive, and triggers decoherence and collapse. 
To  \cor{obtain more complete information about the measured quantum} state, one has to repeat the same experiment many times to gain decent statistics and 
 perform tomography. \cor{The latter is a process of quantum state reconstruction using}
a set of incompatible measurements. \cor{It allows one to  recover the density matrix of the corresponding quantum state}.
\cor{However, quantum computations done on contemporary quantum devices are often subjected to imperfections and noise. While many mitigation and error correction tools have been proposed \cite{mit1,mit2,mit3},
they all assume that the working Hilbert space is restricted. It is therefore of utmost importance to ensure the dimension of the working Hilbert space satisfies such assumptions before applying those techniques.}	

Although the number of outcomes is always quite arbitrary, \cor{it is impossible for} the measurement \cor{ to} increase the dimension of the input states. \cor{This means that if the working Hilbert space is out of the assumed bounds, it must have been so during the final quantum space preparation.}
There are methods that determine the dimension of the space of a single state. They \cor{most often make use of} a dimension witness, usually in the prepare-and-measure (PM)
scenario, i.e. one prepares \cor{the state and the measurement, each selected from a predefined set}. The matrix of probabilities obtained \cor{in such test}
is constrained by the dimension \cor{of the working Hilbert space}. The early \cor{dimension} witnesses, based on inequality \cite{gallego,dim1,expdim}
\cor{are useful to confirm the lower bound on the dimension, but when testing upper dimension bound, they}
are outperformed by the \cor{so-called} null witnesses. \cor{In the latter, } the equality to zero is a signature of the dimension \cite{dim,chen,opt,bb22}. Null witnesses turn out to be extremely precise in the detection of an extra space in \cor{ contemporary quantum} devices \cite{epja,aqt}.

The situation becomes more complicated if the measured state itself comes from two parties, especially when they are independent of each other. \cor{
In such cases, the working Hilbert space dimension can be checked using} the Schmidt number of a bipartite state \cite{sch1,sch2}, i.e. the locally irreducible dimension of the \cor{given} state.
It allows replacing the inefficient PM scenario by the measure-and-measure (MM) protocol, when only measurements are chosen \cor{by} each party,
or even \cor{where} one makes just a single measurement with many outcomes. Such tests turned out to be extremely precise when quantifying the
entanglement space \cite{epjb}.

Here we propose yet another perspective, namely, to test the dimension of two parties in a joint measurement, i.e. prepare-and-prepare (PP) scenario.
\cor{ It can be thought of as} reversing the MM setup, with a few caveats. \cor{In the PP scenario, }each party chooses to prepare its state independently in one of the \cor{predefined} initial states. 
\cor{It is critical to assume independence between parties, as otherwise the additional correlations may affect the conclusion.} 
The measurement channel does not depend on the states, but its outcome - resulting  in a probability matrix, whose rank corresponds to the bipartite dimension \cor{- does}. The  witness \cor{, in this scenario,} is the \cor{probability matrix determinant}\cite{bell-det}. \cor{ It will be} 
equal to zero for sufficiently large sets of prepared states, with finite statistical error. 

The test is robust against local imperfections, such as gate errors, \cor{as long as they are} within the two-level space. \cor{It's also free of the issues that PM and MM scenarios have.}
In contrast to PM scenario, \cor{in PP scenario} the operations on qubits are performed in parallel \cor{(not sequentially), which reduces the possibility of their mutual influence}. Due to joint measurement, \cor{the issue of measurements correlations, present in MM scenarios, vanishes}. 

We have demonstrated the validity of the test  on IBM Quantum. \cor{Our experiment included two parties sharing a qubit. Aside from the shared qubit, each party had a single qubit at their disposal. 
They prepared one of the predefined single-qubit states, independently. After that, the final state of the shared qubit was prepared. The test concluded in the shared qubit state measurement. } Most tests were positive, especially \cor{on} the newest IBM Heron devices.
However, there are several strong failures. They cannot be explained by simple technical imperfections, as the test
is robust to local noise and the quantum operations \cor{we use} differ only by a phase, and \cor{are } not synchronized with the other qubits. 

The paper is organized as follows. We start from the witness construction. Then we show the implementation of the test on IBM Quantum \cor{platform. Finally we present and discuss the results}, paying \cor{close} attention to their reliability and \cor{to the tested devices} calibration errors. 
Detailed calculations are given in the Appendices.

\begin{figure}
\includegraphics[scale=.3]{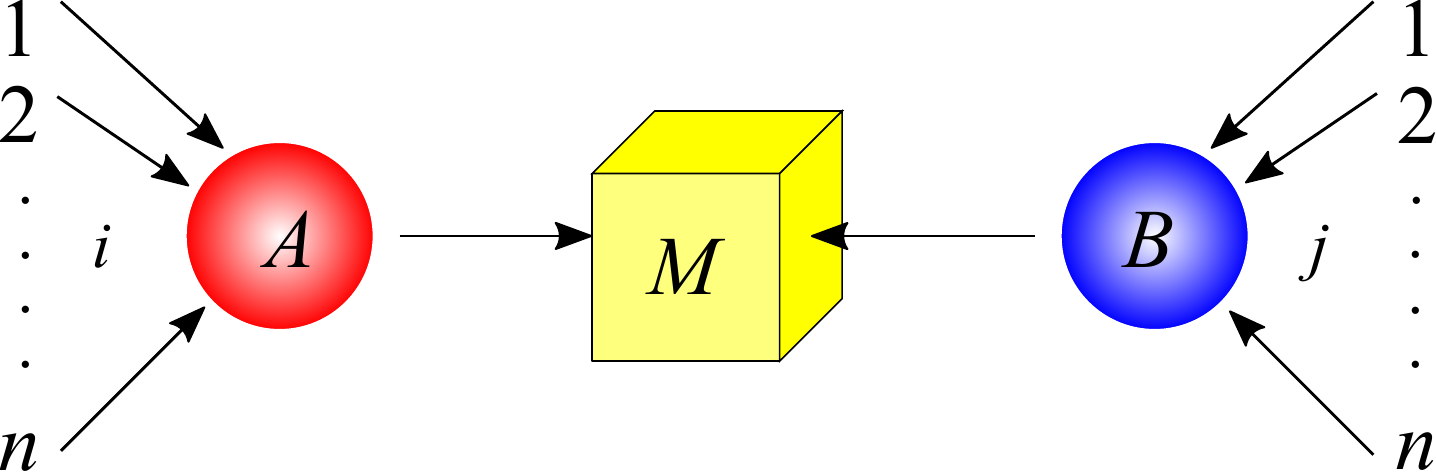}
\caption{Two parties $A$ and $B$ perform either one of $n$ independent preparations, indexed by $i$ and $j$, respectively. The joint measurement takes the prepared states as inputs.}
\label{mab}
\end{figure}

\begin{figure}
\includegraphics[scale=.3]{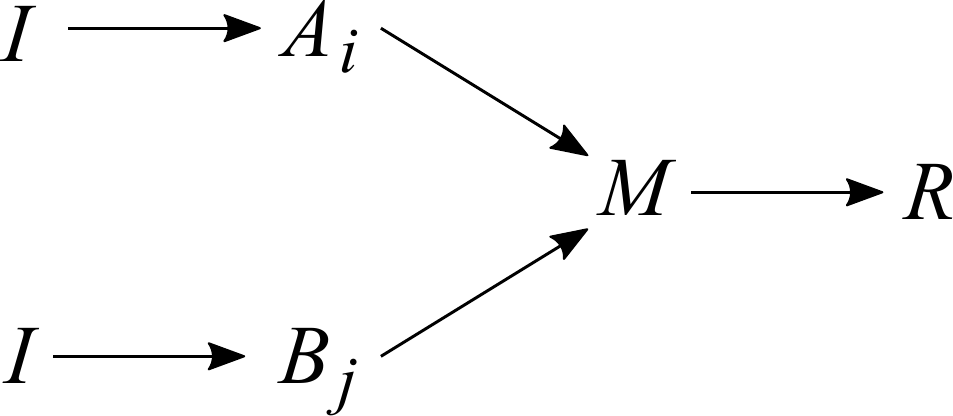}
\caption{The \cor{operation flowchart of the prepare-and-prepare scenario} protocol. Two independent systems are initialized ($I$),
and independently prepared by respective operations $A_i$ and $B_j$. Then a joint measurement $M$ is applied, and the protocol is closed by readout $R$ (dichotomic in our proposal).}
\label{mabp}
\end{figure}

\section{Prepare-and-prepare dimension witness construction}
\label{sec:PP}

Suppose we have a composite (tensor) \cor{quantum} system \cor{that consists of two parties - $A$ and $B$. The prepare-and-prepare (PP) scenario begins with the parties independently preparing their local \cor{subsystems}
in one of the $n$ predefined states \cor{$A_i$ and $B_j$}, where $i$ and $j$ index the predefined state. We will refer to those as local states. The local states are Hermitian ($A_i^\dag=A_i$), positive ($A_i\geq 0$), 
and subjected to normalization (via their trace) in the respective subspaces \cite{nielsen}, $i,j=1\dots n$. At this point the state of the considered quantum system is} $\rho_{ij}=A_i\otimes B_j$. To shorten the notation, we drop the tensor sign 
whenever \cor{it is} unambiguous, i.e. $AB\equiv A\otimes B$. \cor{The  protocol is then concluded} by the joint measurement $\mathcal M$ in the combined space, such that $0\leq \mathcal M\leq 1$. \cor{In our proposition $\mathcal M$ has} dichotomic outcome, 
e.g. \cor{it outputs} $0$ or $1$. \cor{For any selected outcome $s$, the probability of $s$, for given $A_i$ and $B_j$, will read:
\be
p^s_{ij} \equiv p_{ij}=\mathrm{Tr}\mathcal M_s A_iB_j\label{pij},
\ee
where $\mathcal M_s$ is the measurement operator corresponding to outcome $s$.}
The \cor{flowchart} of the protocol is depicted in Fig. \ref{mabp}.

\cor{We can now define} $p$ \cor{as } a $n\times n$ matrix of probabilities $p_{ij}$ for all combinations of states prepared independently by $A$ and $B$.
The rank \cor{of} $p$ depends on the \cor{local systems dimensions - $d_A$ and $d_B$ for $A$ and $B$ systems, respectively - and is} loosely related to the old definition of the Schmidt number \cite{sch1,sch2}. \cor{From now on, we will use the term} \textit{Schmidt number} \cor{to address $d = \min(d_A, d_B)$.}

\cor{The dependence between rank of $p$ and the minimal dimension $d$ allows us to construct null $d$ dimension witness assuming}
\begin{itemize}
\item initial independence of the systems $A$ and $B$,
\item no mutual influence between the systems during the preparations,
\item fixed minimal expected dimension \cor{$d_{min}$} in real or quantum space.
\end{itemize}
\cor{The witness $W_n$ will be the determinant of $p$, $W_n=\det p$. For adequately selected set of predefined states (see section \ref{sec:test}), the witness value is
expected to be} zero (up to statistical error) if the above assumptions are satisfied and can be nonzero otherwise. \cor{This means $W_n=0$} if the Schmidt number satisfies 
\begin{itemize}
	\item $d<n$ (classical case),
	\item $d^2<n$ (quantum complex case), 
	\item $d(d+1)/2< n$ (quantum real case).
\end{itemize}
\cor{That is} because the size of $p$ exceeds the maximal rank of the set of allowed matrices $A_i$ or $B_j$, which spans the available linear space.
If the \cor{observables linear space dimension} is smaller than the size of $p$, then some observable must be a linear combination of the rest.
By linearity of $p$ (eq. \ref{pij}) as a function of the observables, the same applies to its columns and rows, \cor{ thus} the determinant must vanish. In particular, in the case \cor{$d_{min}=2$}, the witness is zero for $n>2$ in the classical case, $n>3$ in the quantum real case, and $n>4$ in the full complex case. If $W_5\neq 0$ then we have a classical system of dimension 5 or a quantum one of dimension 3 (real or complex). In the non-zero cases, $W_n$ is still bounded. The classical maximum is reached whenever $d=n$, by the Hadamard determinant, i.e.
$A_i=|i\rangle\langle i|=B_i$ with \cor{$\mathcal M=m_{ij}|ij\rangle\langle ij|$}. The \cor{maximal} values of \cor{$W_n$ for different $n$} are given in Table \ref{tabone} and the corresponding matrices \cor{$m$} are shown in Table \ref{tabtwo}.

\begin{table}
\begin{tabular}{*{10}{c}}
\toprule
$n$&1&2&3&4&5&6&7&8&9\\
\midrule
$\mathrm{max}\:W_n$&1&1&2&3&5&9&32&56&144\\
\bottomrule
\end{tabular}
\caption{The classical maximum of $W_n$, equivalent to a maximal determinant of the $n\times n$ matrix with entries of $0$ or $1$.}
\label{tabone}
\end{table}

\begin{table*}
$$
\begin{pmatrix}
1\end{pmatrix},\:
\begin{pmatrix}
1&0\\
0&1\end{pmatrix},\:
\begin{pmatrix}
1&0&1\\
1&1&0\\
0&1&1\end{pmatrix},\:
\begin{pmatrix}
0&1&1&1\\
1&1&0&1\\
1&0&1&1\\
1&1&1&0
\end{pmatrix},\:
\begin{pmatrix}
1&0&0&1&1\\
0&1&0&1&1\\
0&0&1&1&1\\
1&1&1&0&1\\
1&1&1&1&0\end{pmatrix},\:
\begin{pmatrix}
1&0&0&1&1&0\\
0&1&0&0&1&1\\
1&0&1&0&0&1\\
1&1&0&1&0&0\\
0&1&1&0&1&0\\
0&0&1&1&0&1\end{pmatrix},$$
$$
\begin{pmatrix}
1&0&1&0&1&0&1\\
0&1&1&0&0&1&1\\
1&1&0&0&1&1&0\\
0&0&0&1&1&1&1\\
1&0&1&1&0&1&0\\
1&1&0&1&0&0&1\\
0&1&1&1&1&0&0
\end{pmatrix},\:
\begin{pmatrix}
1&0&1&0&0&1&1&0\\
1&1&0&1&0&0&1&1\\
1&1&1&0&1&0&0&1\\
0&1&1&1&0&1&0&0\\
0&0&1&1&1&0&1&0\\
1&0&0&1&1&1&0&1\\
0&1&0&0&1&1&1&0\\
0&0&1&0&0&1&1&1
\end{pmatrix},\:
\begin{pmatrix}
0&1&1&1&1&1&1&0&0\\
1&0&1&1&1&1&0&1&0\\
1&1&0&1&1&0&1&1&0\\
1&1&1&0&0&1&1&1&0\\
1&1&1&0&1&0&0&0&1\\
1&1&0&1&0&1&0&0&1\\
1&0&1&1&0&0&1&0&1\\
0&1&1&1&0&0&0&1&1\\
0&0&0&0&1&1&1&1&1
\end{pmatrix}
$$
\caption{Binary matrices $n\times n$ with the maximal $W_n$
given in Table \ref{tabone}.}
\label{tabtwo}
\end{table*}

\begin{table}
\begin{tabular}{c|*{5}{c}}
\toprule
$n\backslash d$&$2r$&$2c$&$3r$&$3c$&$4rc$\\
\midrule
$2$&$1$&&&&\\
$3$&$0.316$&&$2$&&\\
$4$&$0$&$0.111$&$0.903$&&$3$\\
$5$&$0$&$0$&$0.297$&$0.333$&$1.874$\\
\bottomrule
\end{tabular}
\caption{The calculated fully quantum maxima for the test of the witness quantity $W_n$ for the dimension $d$, with $(r/c)$ denoting the real/complex case.
The empty cell takes the nearest value on the left.}
\label{tbq}
\end{table}

\section{The test}
\label{sec:test}

\begin{figure}
	\includegraphics[scale=.5]{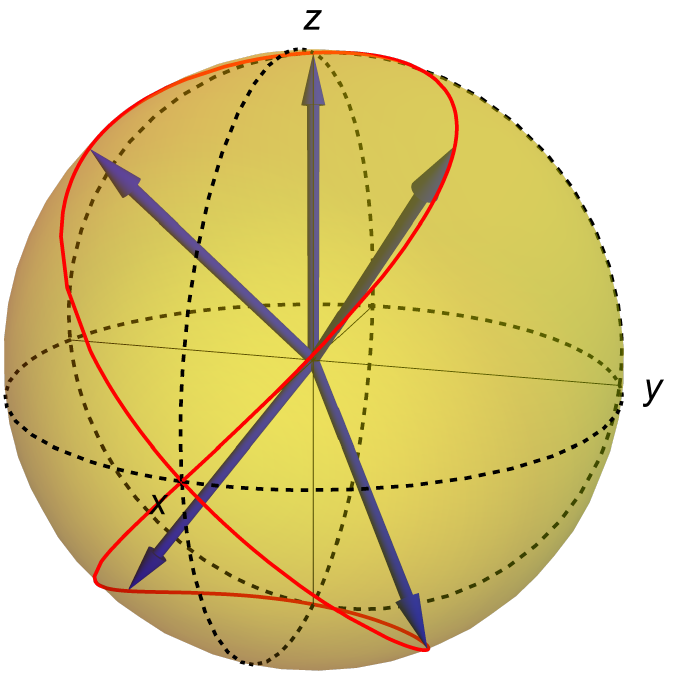}
	\caption{The Bloch vectors \cor{representing predefined states} used in this work. For the test of \cor{total} qubit dimension, the vectors should span the in all three dimensions. Due to the IBM implementation of  single-qubit rotations, the simplest realization of this requirement, with  only-phase-dependent gate, are vectors on the Viviani curve, corresponding to the states $S_\alpha S|0\rangle$ (see section \ref{sec:test}).}
	\label{bloch}
\end{figure}

\cor{Using the null witness defined in section \ref{sec:PP}, we can define the discrimination test for the dimension of the bipartite system space. The set of local one-qubit states available to each party, is the subset of states lying } on the non-planar Viviani curve on the Bloch sphere. \cor{Let us recall that the Viviani curve arises as the 	intersection of a cylinder tangent to a sphere and embedded inside it, passing through the origin (see figure \ref{bloch})}. \cor{While other choices might be possible, we would like to notice that similar approach, but using planar curve, would not be possible. Such } choice automatically reduces the dimension \cor{of the tested space, influencing the results of the test. Selection of the parametric Viviani curve is also convenient, as all the states lying on such curve are accessible from $\ket{0}$ using only $\pi/2$ rotations by $S$ gate and phase-shifted rotations $S_\alpha$.} 

For the sake of clarity, let us focus on party $A$. \cor{We will} use the Bloch sphere representation of \cor{ the} states. \cor{In this form, we use} vectors $\boldsymbol a$ to represent the state $A=|\boldsymbol a\rangle\langle \boldsymbol a|=(1+\boldsymbol a\cdot\boldsymbol \sigma)/2$,
with Pauli matrices $\sigma_{1,2,3}$.
The initial state $|0\rangle\langle 0|$ corresponds to the vector $(0,0,1)$. \cor{The Viviani curve is given by $\boldsymbol a=-(\sin\alpha\cos\alpha,\sin^2\alpha,\cos\alpha)$}. 
\cor{In the test we will use states in general form} $A_i=|\psi_i\rangle\langle\psi_i|$ with
$|\psi_i\rangle=S_i X_+|0\rangle$ for $S_i\equiv S_{\alpha_i}$. \cor{For the test implementation, we selected states corresponding to $\alpha \to \alpha_k=\pi / 4, -\pi / 4, 3\pi/4, -3\pi/4, 0$, on the Viviani curve, for $k = 1,2,3,4,5$, respectively.} 
\cor{The gates we used are}
\begin{equation}
	\sqrt{X}=X_+=(I-iX)/\sqrt{2}\label{smat},
\end{equation}
denoting Pauli gates $I=\sigma_0=1$, $X=\sigma_1$, $Y=\sigma_2$, $Z=\sigma_3$,
in the $|0\rangle$, $|1\rangle$ basis (see the notation of all gates we 
use in Appendix \ref{appb}).
The rotation for a given angle $\theta$ is realized with the native gate $X_+$ and two gates $Z_\theta$
\begin{equation}
	S_\theta=Z^\dag_\theta X_+Z_\theta
\end{equation}
with
\begin{equation}
	Z_\theta=I\cos\theta/2-iZ\sin\theta/2,
	\label{sgate}
\end{equation}
and the shorthand notation $Z=Z_{\pi}$, $Z_\pm=Z_{\pm\pi/2}$.

\begin{figure*}
	\begin{tikzpicture}[scale=1]
		\begin{yquant*}[register/separation=3mm]
			qubit {} q[3];
			[name=init]
			init {A $\ket 0$} q[0];
			init {M $\ket 0$} q[1];
			init {B $\ket 0$} q[2];
			box {$X_+$} q[0];
			box {$X_+$} q[2];
			box {$S_i $} q[0];
			box {$S_j $} q[2];
			cnot q[1] | q[0];
			cnot q[0] | q[1];
			[name=luA]
			cnot q[2] | q[1];
			box {$X_+$} q[1];
			box {$Z_{\pi/4}$} q[1];
			box {$S$} q[1];
			cnot q[1] | q[2];
			[name=rdA]
			box {$Z$} q[1];
			box {$X_+$} q[1];
			box {$Z_{-\pi/4}$} q[1];
			[name=ruA]
			box {$S$} q[1];
			measure q[1];
			\node[draw, dashed, fit=(luA) (luA) (ruA) (rdA), "$\mathcal M\to \mathcal M'$" below] {};
		\end{yquant*}
	\end{tikzpicture}
	\caption{The \cor{ quantum circuit} of the test \cor{ as implemented in our experiments on} IBM Quantum \cor{devices}. \cor{
We have framed the unitary transformation of the projection $\mathcal M$ from (\ref{mmo}) to the directly measured state $\mathcal M'$.
Notice that the circuit run on the actual machine differs from the schematic, as it required the use of the device-native gates. See appendix \ref{appb} for details. The $CNOT$ gates for $|ab\rangle$ (linked) have the control qubit $a$ and
the target qubit state $b$, $CNOT_\downarrow|ab\rangle=|a\;a\oplus b\rangle$,  depicted as $\bullet$ and $\oplus$. respectively.}}
	\label{mid23}
\end{figure*}

\cor{After the preparation stage we are ready to make a measurement.
The necessary physical operation must entangle the qubits, e.g. $CNOT_\downarrow|ab\rangle=|a\; b\oplus a\rangle$
and $CNOT_\uparrow|ab\rangle=|a\oplus b\; b\rangle$ with $a\oplus b$ meaning addition modulo 2 (see Appendix \ref{appb}).
It it tempting to take any joint measurement that takes input from both states, e.g. projection onto the Bell entangled state
\be
\sqrt{2}|\psi\rangle=|00\rangle+|11\rangle.
\ee
Despite formal simplicity and implementation by a single entangling gate (e.g. $CNOT$), this choice is sensitive to errors.
\cor{ Such measurements} may be correlated with the \cor{state} preparation. \cor{To bypass this shortcoming, } we decided to apply a bit more complicated measurement protocol, in which
\begin{itemize}
\item only a single qubit is measured,
\item the measured qubit is different from $A$ and $B$.
\end{itemize}
The \cor{ancillary} qubit $M$ acts as a referee, independent from potential bias from qubits $A$ and $B$.}

The test can be implemented as follows. \cor{We start with arbitrarily selecting
one of the qubit $M$ measurement outcomes. The measurement of $M$ is dichotomic (outcomes $0$ and $1$). From now on, we assume we selected 0 as the outcome used to construct matrix $p$. We can now prepare the circuits.}
  
\cor{The circuit begins with the initialization of $A, M$ and $B$ qubit states. $M$ state is set to $\ket{0}$ while states of $A$ and $B$ are selected from the predefined set of states. We then} swap the states of $A$ \cor{and} $M$  \cor{qubits} by a pair of $CNOT_\uparrow CNOT_\downarrow$ on \cor{system} $AM$. \cor{This makes the final state of $A$ qubit the same as the initial state of $M$ i. e. $|0\rangle$.}

\cor{
Now we want to consider the measurement. Usually, all technical realizations of the measurements are in  diagonal $|0\rangle$, $|1\rangle$ eigenbasis. Our measurement of the two-qubit state $MB$ (formerly $AB$) must therefore be mapped on the measurement basis of $MB$ into a projection 
$\mathcal M'=|00\rangle\langle 00|+|01\rangle\langle 01|$.
Note that direct application of $\mathcal M'$ is incorrect, as it would disregard the qubit $B$ completely. The final measurement must include the
information from the $B$ state. In what follows, we will show that
the actual measurement operator that correctly realizes our needs is
\be
\mathcal M_{M\to0} \equiv \mathcal M=|00\rangle\langle 00|+(|01\rangle+|10\rangle)(\langle 01|+\langle 10|)/2. \label{mmo}
\ee
It is also rank 2 projection. {What we need now is an unitary transformation $\mathcal M\to\mathcal M'$.} The mapping $\mathcal M\to \mathcal M'$ is depicted in Fig. \ref{mid23}.}

	\cor{Since technically we can only perform the projective measurement onto the $\ket{0}, \ket{1}$ eigenbasis, we need to implement the $\mathcal M \to \mathcal M'$ mapping using additional gates on $MB$ system. We will show that the subcircuit marked in Fig. \ref{mid23} satisfies the measurement transformation according to our needs. The necessary transforamtion is equivalent to controlled Hadamard gate, $H=(X+Z)/\sqrt{2}$, since it results in controlled entanglement. Starting from the end of the circuit, the projection of qubit $M$ state on the final state $\langle 0|$ results in an intermediate state of $M$,
	\be
	\langle 0|X_+Z_{-\pi/4}X_+Z=\langle \phi|=i\langle 0|\sin(\pi/8)+i\langle 1|\cos(\pi/8).
	\ee
Assuming that we act on qubits (two-dimensional systems), the final state of $B$ is a superposition of $\bra{0}$ and $\bra{1}$. We have to adjust the state of $M$ to include this information in the final measurement. We will do so using CNOT gates. Notice that $\bra{0}$ part of state $B$ would not influence the state of $M$, as $\bra{\phi 0}  CNOT_\uparrow = \bra{\phi 0}$, and to change $\bra{\phi}$ to the target $\bra{0}$ we only need to apply}
	\be
	\langle\phi|X_+Z_{\pi/4}X_+ = \cor{\langle 0| Z =} \bra{0} ,
	\ee
	since
	\be
	X_+^2=-iX,\:X_+Z=ZX_-,
	\ee
	and again $\langle 00|CNOT_\downarrow=\langle 00|$. \cor{The $\bra{1}$ of state $B$ will however influence the state of $M$ when $CNOT_\uparrow$ is applied to the $MB$ system. Notice that}
	$
	\langle\phi 1|CNOT_\uparrow=\langle\phi' 1|
	$
	with
	$
	\langle\phi'|=\langle \phi|X
	$.
	\cor{The action of this additional $X$ gate manifests in the following way:
	\begin{align}
		&
		\langle\phi'|X_+Z_{\pi/4}X_+=\langle 0|X_+Z_{-\pi/4}X_+ZXX_+Z_{\pi/4}X_+=\nonumber\\
		&\langle 0|X_+Z_{-\pi/4}ZXZ_{\pi/4}X_+=\langle 0|XX_+Z_{\pi/4}ZZ_{\pi/4}X_+=\nonumber\\
		&\langle 1|X_+Z_{-}X_+=(\langle 0|+\langle 1|)/\sqrt{2},
	\end{align}
	ignoring global phases. \cor{With the application of $CNOT_\downarrow$ on the $MB$ system, we're left with}
	\be
	(\langle 01|+\langle 11|)CNOT_\downarrow=\langle 01|+\langle 10|.
	\ee
}\cor{We now see that the subcircuit from figure \ref{mid23} realizes the measurement mapping $\mathcal M \to \mathcal M'$. The mapped measurement completes our protocol and does not affect qubits $A$ and $B$. }

The probability matrix $p$ obtain using this protocol reads
\be
\frac{1}{8}
\begin{pmatrix}
	5-\sqrt{8}&4-\sqrt{8}&4&5&2-\sqrt{2}\\
	4-\sqrt{8}&5-\sqrt{8}&5&4&2-\sqrt{2}\\
	4&5&5+\sqrt{8}&4+\sqrt{8}&2+\sqrt{2}\\
	5&4&4+\sqrt{8}&5+\sqrt{8}&2+\sqrt{2}\\
	2-\sqrt{2}&2-\sqrt{2}&2+\sqrt{2}&2+\sqrt{2}&0
\end{pmatrix},
\ee
with its adjugate \cor{(see section \ref{sec:err})}
\be
\frac{1}{2^9}\begin{pmatrix}
	-1&+1&-1&+1&0\\
	+1&-1&+1&-1&0\\
	-1&+1&-1&+1&0\\
	+1&-1&+1&-1&0\\
	0&0&0&0&0\end{pmatrix}.
\ee

\cor{ 
Apart from the general assumptions, the test is robust against
\begin{itemize}
	\item arbitrary initial states, as long as they remain independent;
	\item arbitrary local operations;
	\item interaction between the middle detector $M$ and the parties $A$ and $B$, confined to the two level space;
	\item single faulty party, as the test check the lower of the dimensions.
\end{itemize}
If the operation  by two-qubit gates are confined to the $2\times 2$ space, then the test should be passed even if the local
single-qubit \cor{gates experience} leakage to other states.
In other words only serious malfunction, e.g. correlation between initial states, can invalidate the assumption of independence.}

\cor{Notice that t}he test is device-independent. We make no assumption about the actual realization of preparations or measurements. However, we assume independence of $A$ and $B$ as otherwise we cannot write the concomitant tensor product. The test can easily be executed on available quantum computers.

\section{Error analysis}
\label{sec:err}

To \cor{experimentally} determine $W_n$ value, we collect data from $N$ repetitions of each \cor{possible composite system state}. 
The uncertainty in determining $W_n$ is analogous to the prepare-measure scheme in \cite{bb22} 
for finite statistics and assuming $\langle W_n\rangle=0$.
From \cor{the} Laplace expansion, the linear deviation of the determinant for small changes of $p$ reads
\be
\delta W=\sum_{kj}\delta p_{kj}(\mathcal{A})_{jk},\label{lapla}
\ee
where $\mathcal A=\mathrm{Adj}\; p$  is the adjugate matrix (matrix of minors of $p$, without
 row \cor{k} and column \cor{j}, and then transposed). \cor{Here:}
\be
\delta p_{kj}=\bar{p}_{kj}-p_{kj}
\ee
with 
\cor{
\be
\bar{p}_{kj}=N^s_{kj}/N
\ee
}being the actually measured statistics,\cor{ $N^s_{kj}$ is the number of times the selected outcome $s$ measured during the $N$ independent trials}.
By \cor{the independence of the local states preparation combinations}, the variance is the sum over those preparations
\be
\langle (\delta W)^2\rangle=\sum_{kj}{\mathcal A}_{jk}^2\langle(\bar{p}_{kj}-p_{kj})^2\rangle.
\ee
\cor{Moreover}, due to the independence of trials
\be
\langle(\bar{p}_{kj}-p_{kj})^2\rangle=Np_{kj}(1-p_{kj})
\ee
following from the properties of Bernoulli distribution. Finally
\be
N\langle (\delta W)^2\rangle=\sum_{kj}{\mathcal A}^2_{jk}p_{kj}(1-p_{kj}),
\ee
with the adjugate matrix calculated directly, since $p\mathcal A=\det p$ cannot be inverted when $\det p=0$.
One should also avoid the situation of $\mathcal A=0$, i.e. when the rank \cor{of $p$}
is already smaller, as the error \cor{obtained this way} becomes unreliable. \cor{In such cases} one has to consider second-order minors.

\begin{figure*}
\includegraphics[scale=.7]{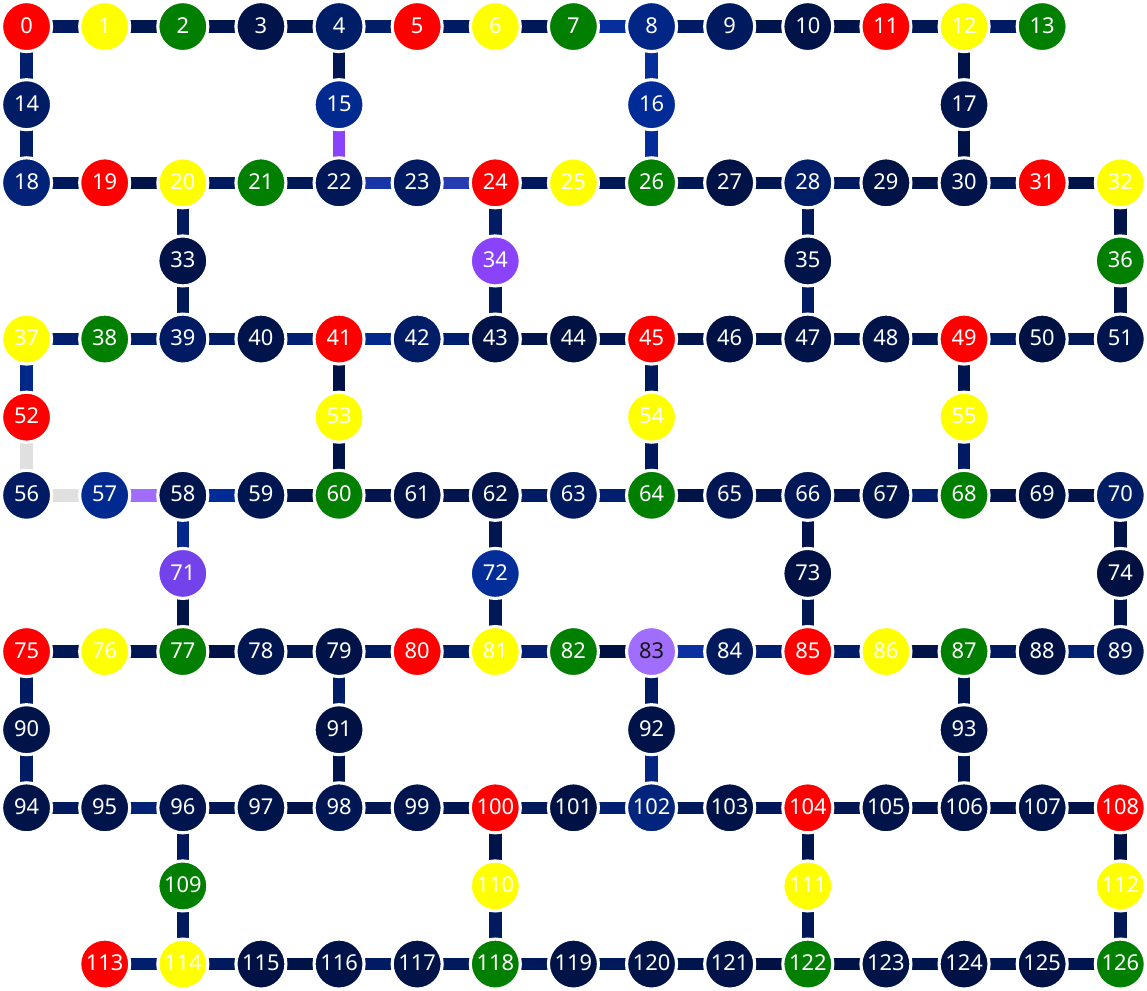}
\caption{Topology of the qubit grid of IBM Quantum devices in Eagle generation, \emph{ibm\_sherbrooke}. The circles represent qubits. \cor{A connection between the qubits indicates a possible direct two-qubit interaction (non-empty set of two-qubit gates) between them}. 
The \cor{actual} grid is hexagonal. \cor{The qubit triplets, selected for the experiment, were marked using following scheme:} 
$A$ -- red, $M$ -- yellow, $B$ -- green.}
\label{top}
\end{figure*}

\begin{figure*}
\includegraphics[scale=.7]{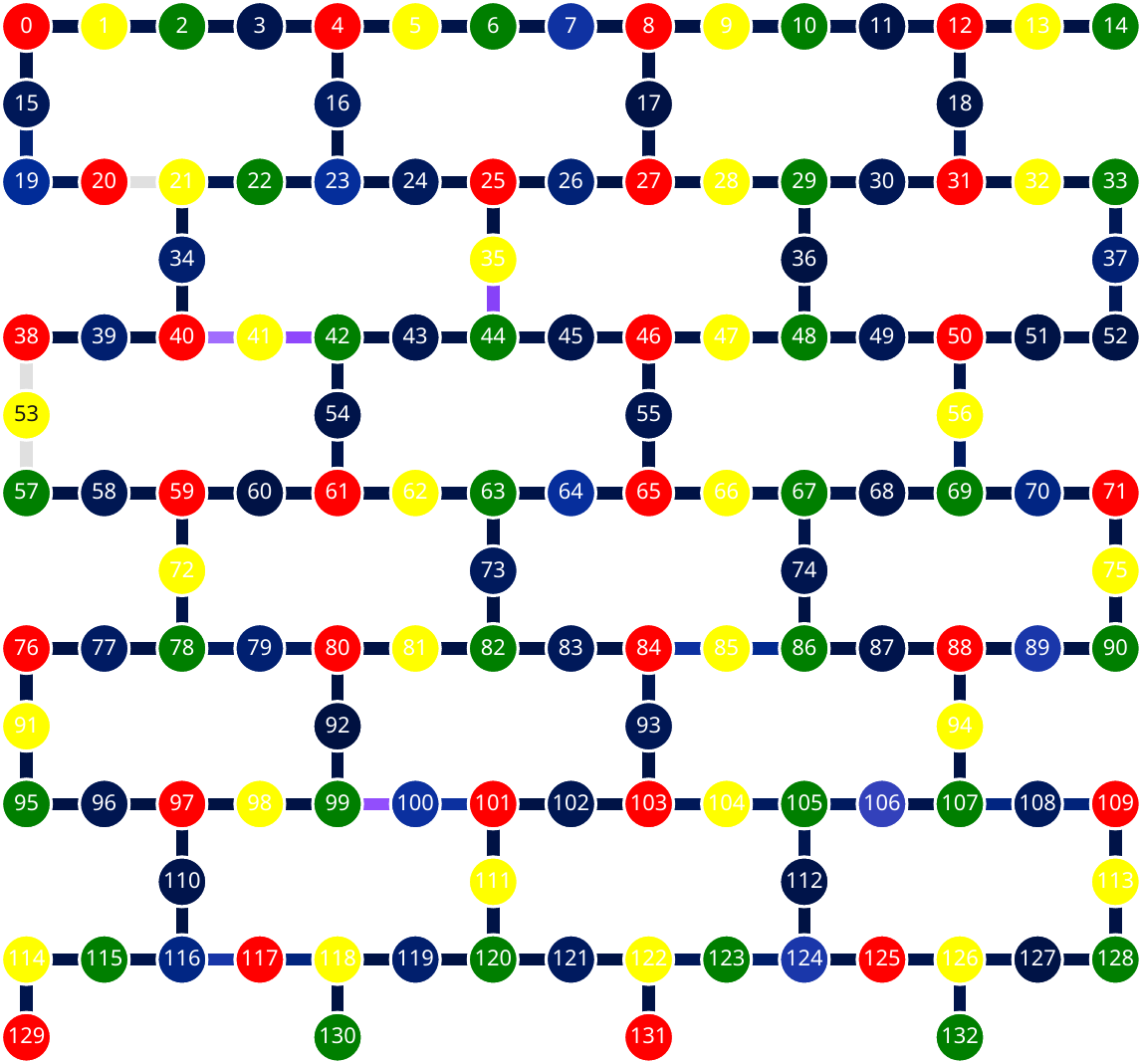}
\caption{\cor{Topology of the qubit grid of IBM Quantum devices in Heron \emph{ibm\_torino}. Notation as in Fig. \ref{top}}}
\label{topt}
\end{figure*}

\begin{figure*}
\includegraphics[scale=.7]{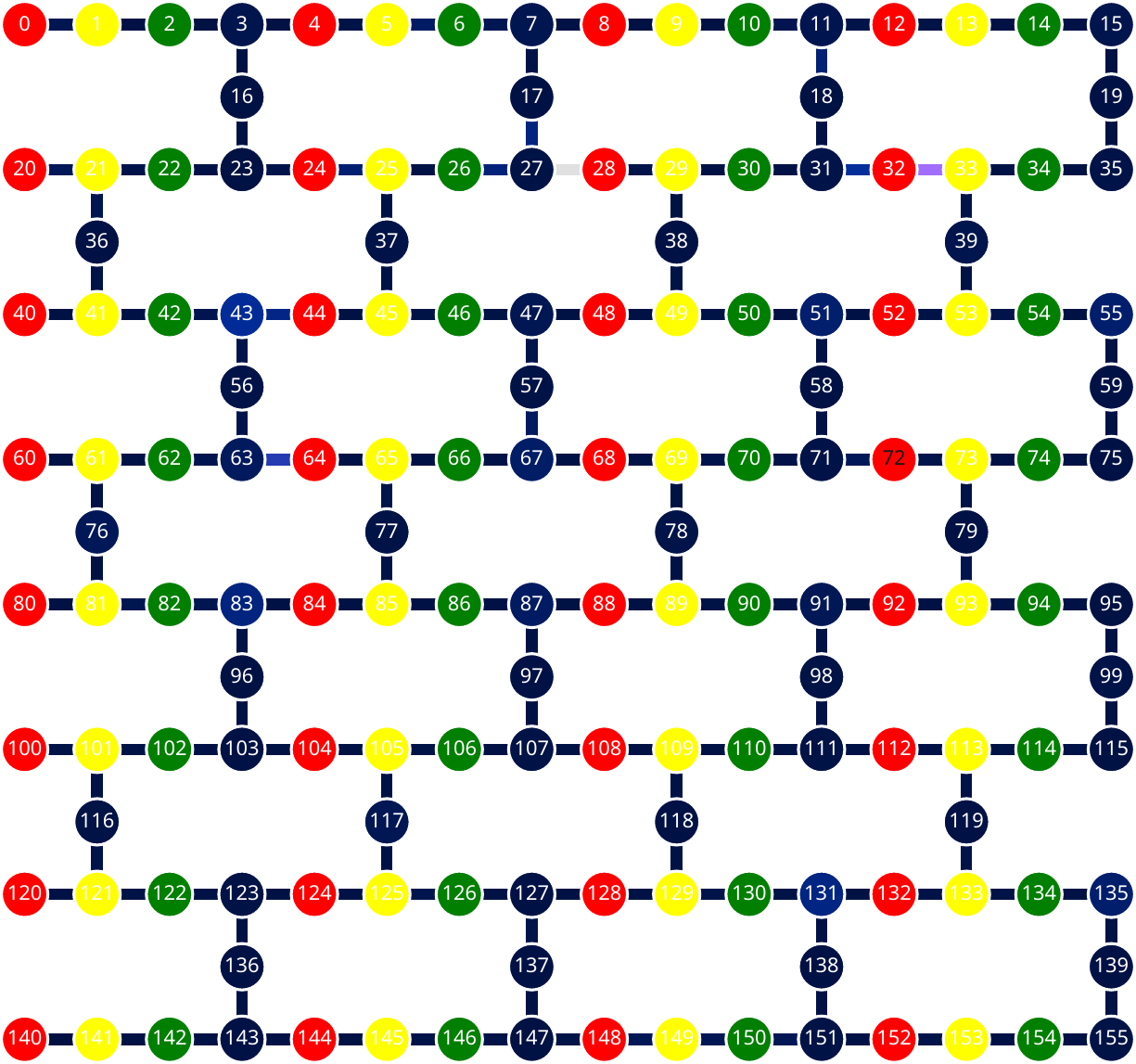}
\caption{\cor{Topology of the qubit grid of IBM Quantum devices in Heron \emph{ibm\_fez} and \emph{ibm\_pittsburgh}. Notation as in Fig. \ref{top}}}
\label{topf}
\end{figure*}

\subsection{Variation of probability}
\cor{The} witness \cor{$W_n$ in our test} is zero only if the probability is identical for all independent trials. 
Suppose there is a small variation of  $p$, e.g. 
\begin{equation}
p_t= \bar{p}+\epsilon\delta p_t
\end{equation}
where $t$ is the index of the trial and $\bar{p}$ is the average probability \cor{matrix}
\begin{equation}
\bar{p}=\sum_t p_t/N
\end{equation}
for the total number of trials $N$. 
We shall estimate the effect of the small variation parameter $\epsilon\ll 1$ \cor{on the probability matrix}.
We assume that \cor{$W_{n, t}=0$} i.e. the test should pass for each trial independently, but
\cor{ 
\begin{equation}
\bar{W}_n=\det\bar{p}
\end{equation}
}is not necessarily zero.
The effect of the variation can be taken again from Laplace formula (\ref{lapla}),\cor{ 
\begin{equation}
W_{n, t}=\det(\bar{p}+\epsilon\delta p_t)\simeq \det\bar{p}+\sum_{kj}\epsilon\delta p_{t kj}\bar{\mathcal{A}}_{jk}
\end{equation}
}
As \cor{$W_{n, \alpha}=0$} by definition, we have
\cor{ 
\begin{equation}
N\bar{W}_n \simeq-\epsilon\sum_{t kj}\delta p_{t kj}\bar{\mathcal{A}}_{jk}
\end{equation}
}However summing over all trials, the right hand side will be zero as $\sum_t p_{t kj}=0$ by definition.
We conclude that
\begin{equation}
\bar{W}\propto \epsilon^2
\end{equation}
with the coefficient depending on second-order minors of $\bar{p}$. \cor{It means} that instabilities can affect the witness in the second power of the variation.
This observation is important as only long timescale drifts can be captured by dividing the data
into bins and finding $W_n$ separately. Short time variation can affect the results, but only if it is sufficiently large to give considerable second-order correction.

\section{Demonstration on IBM Quantum}

We have demonstrated the feasibility of the above test on \cor{selected contemporary} IBM Quantum devices.
According to IBM documentation, the gates work in the specified part of the Hilbert space, two-level for a single qubit, and \cor{four-level} for two-qubit gates \cite{ibmdoc}. \cor{In principle, }each gate is isolated from the others. The experiment on such device is completed with the measurement (readout), which
is registered as a random dichotomic variable, $0$ or $1$. \cor{ This makes IBM devices capable of satisfying our witness assumptions.} We stress that IBM does not report any correlations between initial states of different qubits.

Physically the gates are implemented as a manipulation on transmons as qubits with the inter-level drive frequency at the microwave level.
A microwave pulse tuned to the inter-level drive frequency allows one to apply the parametric controlled gates. The native \cor{IBM Quantum} single-qubit gate is the $\pi/2$ rotation \cor{over $X$ axis with additional phase shift. This gate is usually called $\sqrt{X}=X_+$}. 
In addition, there is a native two-qubit  gate on each IBM quantum device, \cor{ which allow us to realize the $CNOT$ operation}.

The \cor{Eagle family devices (\emph{ibm\_sherbrooke}, \emph{ibm\_brisbane})} use Echoed Crossed Resonance ($ECR$) gate, instead of $CNOT$ but one can transpile
the latter by additional single-qubit gates. 
The newest Heron family (\emph{ibm\_torino}, \emph{ibm\_fez}, \emph{ibm\_pitsburgh}) uses CZ (Controlled $Z$) which is symmetric with respect to qubits and transpile easily to $CNOT$. \cor{We included all transpilation details in Appendix \ref{appb}.}.

Each test consists of a certain number of jobs, \cor{so-called in the \texttt{qiskit} environment}. \cor{Within a single job, one can define multiple quantum circuits. Single circuit execution is called shot. The number of shots is specified by the user for all the circuits within a job. This mean that the total number $N$ of each circuit execution is $N=\#jobs\#shots\#repetitions$.} 

\cor{Since the access to IBM Quantum devices is limited, it is important to use
the device time to its fullest. Since IBM limits only the number of total executions within a single job (the limit is $10^7$ shots \cite{ibmdoc}), and we only have 25 distinct experimental setups (quantum circuits), we could repeat the same circuits within a given job.}

Due to calibration changes, \cor{that occur} every several hours, the \cor{measured} probabilities may drift. \cor{This }can affect the witness, being a nonlinear function of probabilities. To take \cor{this effect} into account, we have calculated the witness in two ways \cite{ibm}:\cor{ 
\begin{itemize}
	\item  $W_n$, $\Delta W_n$ -- the witness is obtained from total probabilities of all jobs together;
	\item $W'_n$, $\Delta W'_n$ -- the witness is obtained by calculating probabilities and the witness for each job individually, and then averaging it over jobs.
\end{itemize}
} It turned out that \cor{the witness} values indeed differed, but did not change the verdict about the dimension.
We have run two types of gates scheduling: \emph{as late as possible} (ALAP or L), \emph{as soon as possible} (ASAP or S).
They differ by the timing of the gate operations. In the ALAP scheduling, the gates are applied at the very last to avoid premature decoherence of the qubits, which are initially
in the ground state $|0\rangle$. In the ASAP scheduling, the gates are applied immediately after the start of the shot or after the previous operation.

\begin{figure*}
\includegraphics[scale=.7]{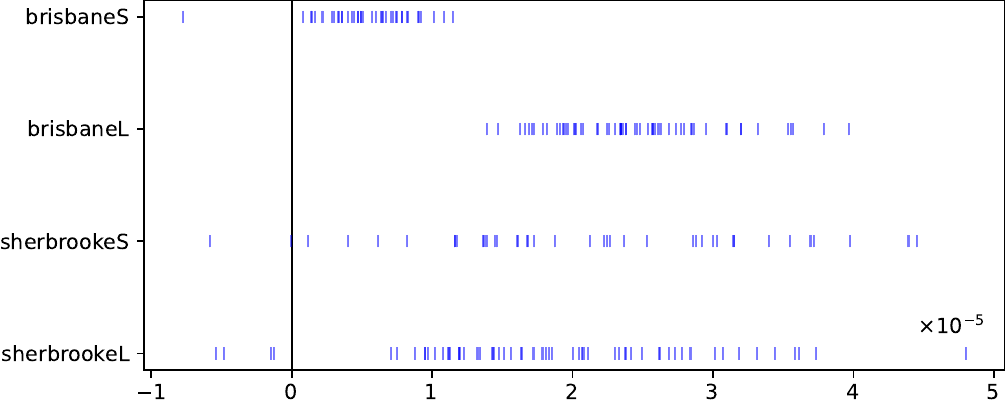}
\caption{The individual experimental values of the witness for the most faulty sets in \emph{ibm\_brisbane} and \emph{ibm\_sherbrooke}, for ALAP/ASAP, tests,
brisbaneL/S, sherbrookeL/S, respectively.}
\label{scat}
\end{figure*}

\begin{figure*}
\includegraphics[scale=.7]{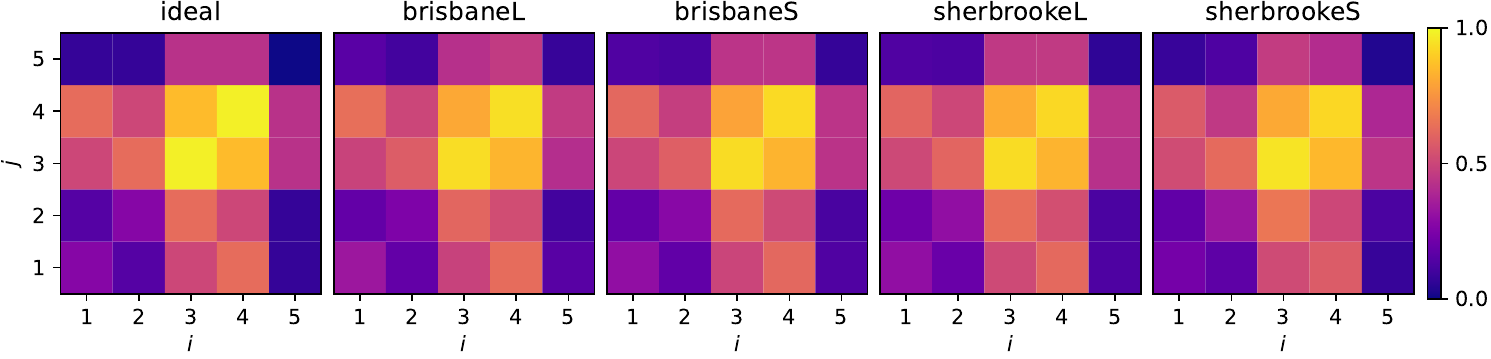}
\caption{The experimental probabilities $p_{ij}$ for the most faulty sets in \emph{ibm\_brisbane} and \emph{ibm\_sherbrooke}, for ALAP/ASAP, tests,
brisbaneL/S, sherbrookeL/S, respectively, \cor{ along} the ideal prediction.}
\label{mesh}
\end{figure*}

\cor{For } ALAP/ASAP scheduling we have run 60/40 jobs, with 4 repetitions and 20000 shots each, \cor{on Eagle devices}. \cor{In each job, the order in which we executed the circuits was random, to avoid memory-related issues.}
Out of 17 preselected sets of qubits, shown in Fig. \ref{top}, on \emph{ibm\_sherbrooke} 
and one in \emph{ibm\_brisbane} most of them passed the test. Only two sets, one in \emph{ibm\_sherbrooke} and one in \emph{ibm\_brisbane}\cor{, failed}. 
The final witness for these cases disagrees with the null hypothesis for $d=2$ beyond 10 standard deviations see Table \ref{resu}. 
However, \cor{in the case of \emph{ibm\_brisbane} we noticed} 4 times smaller deviation \cor{for ASAP scheduling}. The distribution of witness \cor{values}
for individual jobs is consistent with the average, Fig. \ref{scat}.
The accuracy of our test is especially clear when looking at the probability matrix (see Fig. \ref{mesh}), which naively resembles the theoretical prediction, while the exact calculation of the witness reveals the deviation.

\begin{table*}
\begin{tabular}{*{10}{c}}
\toprule
\emph{device} $A-M-B$ &$W^L_5$&$\Delta W^L_5$ & $W^{L\prime}_5$ & $\Delta W^{L\prime}_5$& $W^S_5$&$\Delta W^S_5$ & $W^{S\prime}_5$ & $\Delta W^{S\prime}_5$& 
$f_{AB}$
\\
\midrule
\emph{ibm\_brisbane} 11-12-13 &24.45& 0.57&24.44& 0.58&5.48& 0.50&5.39& 0.51&-34\\
\emph{ibm\_sherbrooke} 108-112-126&18.33& 1.07&18.43& 1.08&21.23& 1.24&21.00& 1.25&26\\
\bottomrule
\end{tabular}
\caption{The experimental values of the witnesses $W$ and $W'$ with their errors $\Delta W$ and $\Delta W'$ in units $10^{-6}$ for the ALAP ($L$) and ASAP ($S$) tests on 
the most faulty qubits $\{A$,$M$,$B\}$ on Eagle family. We have also shown the difference between frequencies of qubits $A$ and $B$, $f_{AB}=f_A-f_B$
in units MHz.}
\label{resu}
\end{table*}

\begin{figure*}
\includegraphics[scale=.7]{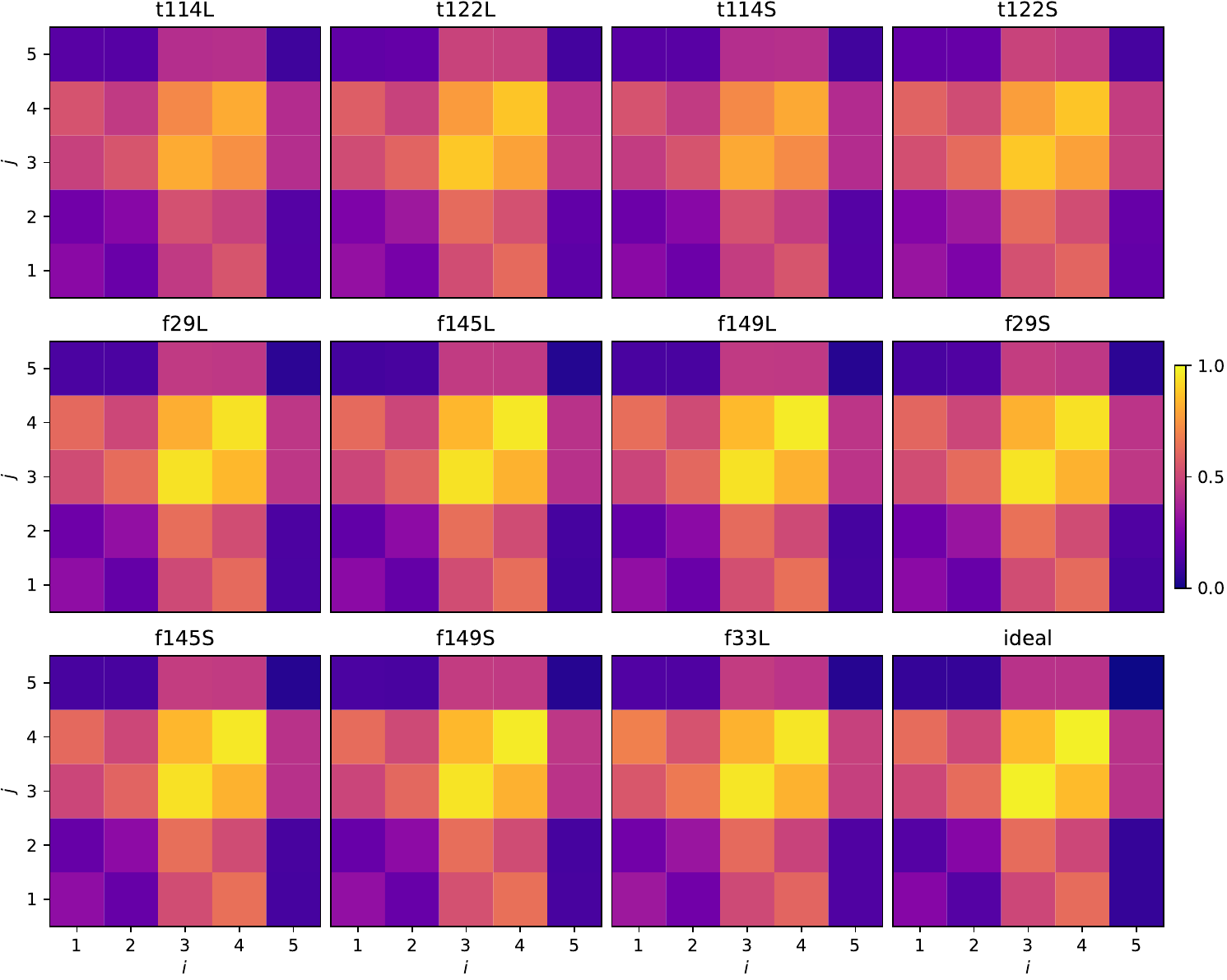}
\caption{The experimental probabilities $p_{ij}$ for the most faulty sets in Heron family with the notation as in Fig. \ref{outh}, \cor{along} the ideal prediction.}
\label{meshh}
\end{figure*}

\begin{figure*}
\includegraphics[scale=.7]{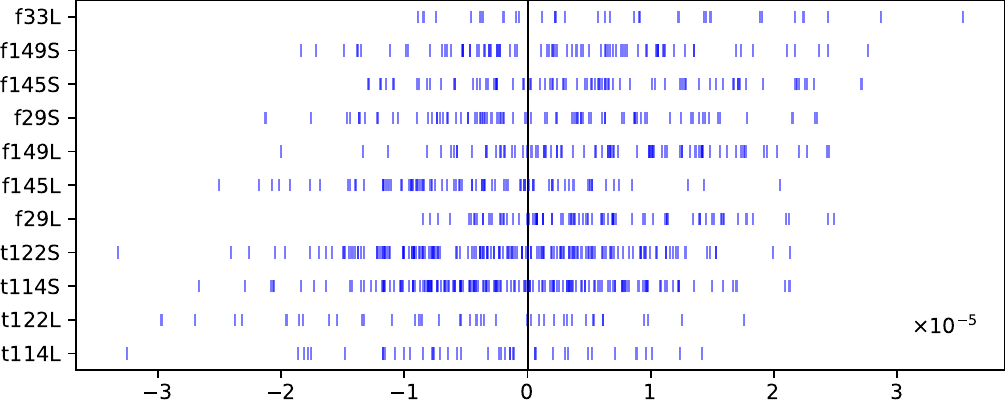}
\caption{The experimental, individual values of the witness for the most faulty sets in Heron family.
The notation: t -- \emph{ibm\_torino}, f -- \emph{ibm\_fez}, p -- \emph{ibm\_pittsburgh}.
the number of the middle qubit $M$ and, 
$L/S$ for ALAP/ASAP, tests,, respectively.}
\label{scath}
\end{figure*}

We have run the tests also on Heron devices, \emph{ibm\_torino}  (40/160 jobs testing 28 sets), \emph{ibm\_fez} (80/80 jobs testing 32 sets), \emph{ibm\_pitsburgh} (30 ALAP only tests on 32 sets). \cor{For \emph{ibm\_torino} and \emph{ibm\_pitsburgh} we selected the triplets 
nonoverlapping and covering the grid uniformly.} The sets are shown on the qubit grid in Fig. \ref{topt} and \ref{topf}.
Most of the qubits have passed the tests i.e. \cor{$W_n$ and $W_n'$} remained within the margin of 5 standard deviations from zero.
We have found still a few outliers and only because of extremely large statistics. Their results are given in Table \ref{outh},
\cor{which also includes} the set of largest nonzero value on \emph{ibm\_pittsburgh} (nearly 5 standard deviations).
The distribution of witnesses for individual jobs is consistent with the average, Fig.~\ref{scath},
and the matrix of probability correspond roughly to the expected pattern, Fig.~\ref{meshh}.
The most striking case is the set 144-145-146 \cor{on \emph{ibm\_fez},} with the significant deviation of opposite sign for the ALAP and ASAP test. The border case on the newest 
\emph{ibm\_pittsburgh} requires more statistics to clarify.

\begin{table*}
\begin{tabular}{*{9}{c}}
\toprule
\emph{device} $A-M-B$ &$W^L_5$&$\Delta W^L_5$ & $W^{L\prime}_5$ & $\Delta W^{L\prime}_5$& $W^S_5$&$\Delta W^S_5$ & $W^{S\prime}_5$ & $\Delta W^{S\prime}_5$
\\
\midrule
\emph{ibm\_torino} 129-114-115 &-3.99& 0.44&-4.15& 0.46&-1.45& 0.22&-1.54& 0.24\\
\emph{ibm\_torino} 131-122-123 &-4.58& 0.68&-5.03& 0.72&-2.19& 0.33&-2.21& 0.34\\
\emph{ibm\_fez}  28-29-30&4.94& 0.91&5.03& 0.91&1.21& 0.89&1.13& 0.89\\
\emph{ibm\_fez}   144-145-146&-4.90& 0.92&-5.01& 0.93&4.83& 0.91&4.71& 0.92\\
\emph{ibm\_fez}   148-149-150&6.32& 0.95&6.32& 0.96&2.44& -0.95&2.71& 0.96\\
\emph{ibm\_pittsburgh} 32-33-34&7.31& 1.67&7.99& 1.76&--&--&--&--\\
\bottomrule
\end{tabular}
\caption{The experimental values of the witnesses $W$ and $W'$ with their errors $\Delta W$ and $\Delta W'$ in units $10^{-6}$ for the ALAP ($L$) and ASAP ($S$) tests on 
the most faulty qubits $A$,$M$,$B$ on Heron family. IBM does not disclose frequencies of Heron devices.}
\label{outh}
\end{table*}

\cor{We have also run noisy simulations using the specifications from} \emph{ibm\_torino}. $W$ and $W'$ remained within the statistical margin from 0,
confirming that there \cor{are} no numerical artifacts.

The data and scripts are publicly available \cite{zen}.

\section{Conclusions and discussion}

We have demonstrated the \cor{prepare-and-prepare} null \cor{witness} test of the Schmidt dimension for the bipartite states. It is complementary to the violation of a Bell-type inequality, \cor{grounded} on similar assumptions, \cor{Prior to its formulation, we discussed the advantages PP scenario has over PM and MM scenarios. }   \cor{A possible extension we envision, could include a dimension witness benchmark that would combine tests using all of the aforementioned scenarios. Such test would be more robust with respect to the assumptions it tests and could} serve as a powerful quality criterion
of multi-qubit networks. \cor{We believe that our test} should help in the diagnostics of quantum devices.

\cor{We would also like to stress} that the null hypothesis tests we presented, remain robust against most common disturbances, as long as they are local,
with known \cor{origin such as environmenal relaxation and dephasing}. Due to the extreme accuracy of the test, we were able to diagnose IBM Quantum devices, far beyond standard technical specifications. The \cor{test} results showed consistency with the Schmidt number $d=2$ in most of the cases. However, occurrence of several outliers is surprising, \cor{ especially considering the declared high quality of the devices}. The deviation is significant and exceeds \cor{typical} origins due to gate errors. The \cor{device shortcomings we report,} require an urgent technical explanation.

\subsection{Possible explanations of deviations}

The general explanations of the nonzero \cor{witness value} can be divided into 3 categories:
	\begin{enumerate}[i)]
		\item extra dimension,
		\item lack of \cor{local states preparation} independence,
		\item variations of probability.
\end{enumerate}

The case i) could be explained by some leakage to other (non-computational states).
To explain the nonzero value of $W$ by small corrections to probabilities of any origin, i.e. $p\to p+\delta p$, one can estimate
$\delta W=\mathrm{Tr} \mathcal A\delta p$ for the adjugate matrix $\mathcal A$.
The elements of the adjugate matrix are of the order $\lesssim 10^{-3}$.
The local leakage to higher excited states is $\lesssim 10^{-6}$ \cite{cay}\cor{. Moreover, considering} that the error is squared, i.e.
$\delta p_{ij}\sim 10^{-12}$, \cor{its total} contribution to the deviation of the determinant \cor{is of order} $\sim 10^{-15}$, \cor{so} much below the observed values $W\gtrsim 10^{-5}$. Nevertheless, even if there is indeed a leakage, then further tests should identify its origin and invent countermeasures.

The case ii) is possible if the operations (gate pulses \cor{applied to} different qubits) are correlated, \cor{e.g. due to frequency proximity} The \cor{$CNOT/ECR/CZ$} gates with errors $\sim 10^{-2}$ could affect the determinant only if they are correlated with \cor{local state} preparations, which \cor{should not ever happen and} would indicate a complete technical failure. Different values of $W$ for ALAP and ASAP tests indicate \cor{that this might actually be the deviations cause}, but the \cor{underlying} mechanism remains unclear.

The probability drifts iii) could be generated e.g. by parity flips \cite{parit}, which change the drive frequency $10^{-4}$ relative to the drive.
Taking our general rule that the effect \cor{of such imperfection} is squared, it gives $W\sim 10^{-8}$, so below the observed deviation. 
Long-time calibration instabilities are excluded by $W\sim W'$. 

In principle none of the suggested explanations can be completely ruled out. \cor{In the end, w}e are unable to present a realistic model explaining the deviations, as the calibration data from IBM do not specify
correlations between initial states, and their magnitude.
It is always possible to invent an artificial model involving e.g. many worlds/copies \cite{plaga,abadp}
($N$ copies of the same system formally boost the dimension from $d$ to $d^N$) but we refrain from going this route yet, as the collected data are
insufficient to draw stronger conclusions. As a positive result, the number of outliers is much smaller in this test compared to prepare-and-measure and measure-and-measure scenarios \cite{epja,epjb,aqt}. Moreover, we also find that the newest IBM Heron family passes the test almost always. We hope that our test will help \cor{in designing future generation of quantum devices.}

\section*{Acknowledgments} The results have been created using IBM Quantum. The views expressed are those of the authors and do not reflect the official policy or position of IBM Quantum team.  \cor{
TR  gratefully acknowledges the funding support by the program "Excellence initiative—research university" for the AGH University of Krakow as well as the ARTIQ project (UMO-2021/01/2/ST6/00004 and ARTIQ/0004/2021).}
We thank Poznañ Supercomputing and Networking Center and ICM UW for the access to IBM Quantum Innovation Center.
We acknowledge technical support of Jakub Tworzyd{\l}o, Bart{\l}omiej Zglinicki, and Bednorz family.

\appendix

\section{Quantum maxima of the witness}
\label{appa}
The following results have been obtained by using a simulated annealing optimization method for the determinant, plus some intuition being guided by the existing symmetry for each case. The ensuing results are later polished by making use of symbolic mathematical packages.

Case $n=3$, $d=2$. Real and complex maximum is $(3/4)^4=0.31640625$ for $A_i=B_i=|\psi_i\rangle\langle\psi_i|$
with
\be
|\psi_j\rangle=\cos(2\pi j)/3|0\rangle+\sin(2\pi j/3)|1\rangle,
\ee
for $j=1,2,3$
and rank $3$ projection
\be
\mathcal M=|00\rangle\langle 00|+|11\rangle\langle 11|+(|01\rangle-|10\rangle)(\langle 01|-\langle 10|)/2.
\label{m2}
\ee
With only two projections, it is lower $1/8=0.125$ in the real case with
\begin{align}
&3\mathcal M=3|00\rangle\langle 00|+3|01\rangle\langle 01|+3|10\rangle\langle 10|\nonumber\\
&-
(|00\rangle+|01\rangle+|10\rangle)(\langle 00|+\langle 01|+\langle 10|),
\end{align}
or $3^6/5^5=0.23328$ in the complex case
with
\be
|\psi_j\rangle=\sqrt{3/5}|0\rangle+\sqrt{2/5}\omega^j|1\rangle,\nonumber\\
\ee
for $ \omega=e^{2\pi i/3}=(\sqrt{3}i-1)/2$,
and 
\be
\mathcal M=|00\rangle\langle 00|+(|01\rangle-|10\rangle)(\langle 01|-\langle 10|)/2.
\ee
For a single projection the maximum is $3^3/2^8=0.10546875$ for both real and complex case
\be
|\psi_j\rangle=\cos(2\pi j/3)|0\rangle+\sin(2\pi j/3)|1\rangle
\ee
for $j=1,2,3$,
and 
\be
\mathcal M=(|01\rangle-|10\rangle)(\langle 01|-\langle 10|)/2.
\ee

Case $n=4$, $d=2$ complex, we get $W=1/9$ for
\begin{align}
&\sqrt{3}|\psi_j\rangle=|0\rangle+\omega^j\sqrt{2}|1\rangle,\nonumber\\
&|\psi_4\rangle=|0\rangle,
\end{align}
for $j=1,2,3$,
and $\mathcal M$ of the form (\ref{m2}).

Case $n=4$, $d=3$ real and complex we obtain the tetrahedral configuration
\begin{align}
&\sqrt{3}|\psi_1\rangle=|0\rangle+|1\rangle+|2\rangle,\nonumber\\
&\sqrt{3}|\psi_2\rangle=|0\rangle+|1\rangle-|2\rangle,\nonumber\\
&\sqrt{3}|\psi_3\rangle=|0\rangle-|1\rangle+|2\rangle,\nonumber\\
&\sqrt{3}|\psi_4\rangle=|0\rangle-|1\rangle-|2\rangle,\nonumber
\end{align}
and rank $5$ projection
\be
\mathcal M=\sum_{i\neq j}(|ij\rangle+|ji\rangle)\langle ij|/2+\sum_i 2|ii\rangle\langle ii|/3-\sum_{i\neq j}|ii\rangle\langle jj|/3,
\ee
for $i,j=0,1,2$,
which gives $-W=2^8 5^4/3^{11}\simeq 0.903204683116282$.

Case $n=5$, $d=3$ real. We numerically retrieve $0.2974087137533708$ for the preparations
\begin{align}
&2|\psi_1\rangle=-|0\rangle+\sqrt{3}|1\rangle,\nonumber\\
&2|\psi_2\rangle=-|0\rangle-\sqrt{3}|1\rangle,\nonumber\\
&|\psi_3\rangle=|0\rangle,\nonumber\\
&2|\psi_4\rangle=-|0\rangle+\sqrt{3}|2\rangle,\nonumber\\
&2|\psi_5\rangle=-|0\rangle-\sqrt{3}|2\rangle,
\end{align}
and  rank $5$ projection
\be
\mathcal M=\sum_{ij}|ij\rangle\langle ij|-\sum_{j=1}^4|\phi_j\rangle\langle\phi_j|,
\ee
for
\begin{align}
&|\phi_1\rangle=|22\rangle,\nonumber\\
&|\phi_2\rangle=p|10\rangle-q|02\rangle,\nonumber\\
&|\phi_3\rangle=p|01\rangle-q|20\rangle,\nonumber\\
&|\phi_4\rangle=x(|12\rangle+|21\rangle)-y|00\rangle,
\end{align}
giving
\begin{align}
&W=(3^8/2^{10})(pq+xy)^2\times\nonumber\\
&(2 x^2 (x^2 + y^2)^2 - 2 p^2 q^2 (x^2+y^2) + p^4) ,
\end{align}
for $p^2+q^2=1=x^2+2y^2$, optimal for
$x^2=0.28105400986117085$, $p^2=0.8688370017274547$.

Case $n=5$, $d=3$ complex. 
We have to take $A_j=|a_j\rangle\langle a_j|$, $B_j=|b_j\rangle\langle b_j|$, with
\begin{align}
&|a_j\rangle=x|0\rangle+y|1\rangle+z\omega^j|2\rangle,\nonumber\\
&|a_4\rangle=s|0\rangle+t|1\rangle,\:|a_5\rangle=|0\rangle\nonumber\\
&|b_j\rangle=p|0\rangle+q\omega^j|1\rangle+r\omega^{2j}|2\rangle,\nonumber\\
&|b_4\rangle=|0\rangle,\:|b_5\rangle=|1\rangle,
\end{align}
with again $\omega=e^{2\pi/3}$,
and  rank $5$ projection
\begin{align}
&\mathcal M=|01\rangle\langle 01|+|02\rangle\langle 02|\nonumber\\
&+
|12\rangle\langle 12|+|13\rangle\langle 13|+|20\rangle\langle 20|+|21\rangle\langle 21|\nonumber\\
&+|\psi_1\rangle\langle\psi_1|-|\psi_2\rangle\langle\psi_2|
-|\psi_3\rangle\langle\psi_3\rangle,
\end{align}
with
\begin{align}
&|\psi_1\rangle=c|00\rangle+d|10\rangle+e|22\rangle,\nonumber\\
&|\psi_2\rangle=f|01\rangle+g|11\rangle+h|20\rangle,\nonumber\\
&|\psi_3\rangle=u|02\rangle+v|12\rangle+w|21\rangle.
\end{align}
It yields
\begin{align}
&W=27t(t\alpha+ 2s\beta) r^2z^2\times\nonumber\\
& (p r e (c x + d  y) - p q h (f  x + g y) - q r w (u x + v y))^2,
\end{align}
with
\begin{align}
&\alpha= 
2 x y u v (c^2 (1 - g^2) - d^2 (1 - f^2)) + \nonumber\\
& (2 x y c d +z^2 (1-h^2))\times\nonumber\\
& ((1 - u^2) (1 - g^2)-(1 - v^2) (1 - f^2))\nonumber\\
&+ (2 x y f g + w^2 z^2) (d^2 (1 - u^2) - c^2 (1 - v^2)) +\nonumber\\
&  z^2 ((c^2 - d^2) (1 - e^2) + (g^2 c^2-d^2f^2) e^2  + d^2 u^2 -  c^2 v^2),
\end{align}
and
\begin{align}
&\beta=y^2(c d (g^2-f^2+u^2(1-g^2)-v^2(1-f^2)) + \nonumber\\
&f g (c^2(1-v^2)-d^2(1-u^2))  + \nonumber\\
&(d^2(1-f^2)-c^2(1-g^2)) u v ) \nonumber\\
&+z^2(u v((1-f^2)(1-h^2) - c^2(1- w^2))\nonumber\\
&-c d ((1-u^2)(1-w^2)-e^2(1-f^2))\nonumber\\
& -f g((1-u^2)(1-h^2)-  c^2 e^2)),
\end{align}
with constraints
\begin{align}
&x^2+y^2+z^2=c^2+d^2+e^2=1,\nonumber\\
&f^2+g^2+h^2=u^2+v^2+w^2=1,\nonumber\\
&p^2+q^2+r^2=s^2+t^2=1.
\end{align}
The last constraint implies the maximum
\begin{align}
&W=27(\alpha/2+\sqrt{\alpha^2/4+\beta^2}) r^2z^2\times\nonumber\\
& (p r e (c x + d  y) - p q h (f  x + g y) - q r w (u x + v y))^2\times\nonumber\\
&(\alpha/2+\sqrt{\alpha^2/4+\beta^2})
\end{align}
which we calculated at $0.33262772907714405$.

Finally, the case $d=4$, $n=5$ gives the maximum for a $5-$cell (a regular simplex in four-dimensional space) configuration, parameterized e.g.
by
\begin{align}
&4|\psi_1\rangle=\sqrt{5}|(|1\rangle+|2\rangle+|3\rangle)-|0\rangle,\nonumber\\
&4|\psi_2\rangle=\sqrt{5}|(|1\rangle-|2\rangle-|3\rangle)-|0\rangle,\nonumber\\
&4|\psi_3\rangle=\sqrt{5}|(|2\rangle-|3\rangle-|1\rangle)-|0\rangle,\nonumber\\
&4|\psi_4\rangle=\sqrt{5}|(|3\rangle-|1\rangle-|2\rangle)-|0\rangle,\nonumber\\
&|\psi_5\rangle=|0\rangle,
\end{align}
with rank $11$ projection
\be
\mathcal M=\sum_{ij}|ij\rangle\langle ij|-\sum_k 16|\tilde{k}\rangle\langle\tilde{k}|/15+4|e\rangle\langle e|/75
\ee
for $|\tilde{k}\rangle=|\psi_k\psi_k\rangle$ and $|e\rangle=\sum_k|\tilde{k}\rangle$.
Here, $\mathcal M$ is essentially a projection orthogonal to each $|\tilde{k}\rangle$.
It gives the matrix entries
\be
p_{ij}=\left\{\begin{array}{ll}
55/64&\mbox{ for }i\neq j,\\
0&\mbox{ for }i=j,
\end{array}\right.
\ee
with the result 
\be
W=4(55/64)^5=1.8748803995549678802490234375.
\ee

\section{Gates and \cor{their transpilation for} IBM Quantum devices}
\label{appb}

\begin{figure}
\begin{tikzpicture}[scale=1]
		\begin{yquantgroup}
			\registers{
			qubit {} q[2];
			}
			\circuit{
			init {$a$} q[0];
			init {$b$} q[1];
			box {$\downarrow$}  (q[0,1]);}
			\equals
			\circuit{
			box {\rotatebox{90}{$CR^+$}}  (q[0,1]);
			box {$X$} q[0];
			box {\rotatebox{90}{$CR^-$}}   (q[0,1]);}
		\end{yquantgroup}
\end{tikzpicture}

\caption{The notation of the $ECR$ gate in the convention $ECR_\downarrow|ab\rangle$.}
\label{ecr}
\end{figure}
We shall use Pauli matrices in the basis $|0\rangle$, $|1\rangle$,
\be
X=\begin{pmatrix}
0&1\\
1&0\end{pmatrix},\:Y=\begin{pmatrix}
0&-i\\
i&0\end{pmatrix},\:Z=\begin{pmatrix}
1&0\\
0&-1\end{pmatrix},\:
I=\begin{pmatrix}
1&0\\
0&1\end{pmatrix}.\label{pauli}
\ee
The IBM Quantum devices  use transmon qubits \cite{transmon}\cor{. On \emph{ibm\_brisbane}} the native single-qubit gates are $X$ and 
\be
X_+=X_{\pi/2}=(I-iX)/\sqrt{2}=\begin{pmatrix}
1&-i\\
-i&1\end{pmatrix}/\sqrt{2},
\ee
denoting $V_\theta=\exp(-i\theta V/2)=\cos(\theta/2)-iV\sin(\theta/2)$ and $V_\pm= V_{\pm \pi/2}$, whenever $V^2$ is identity.
Note that $Z_\theta=\exp(-i\theta Z/2)=\mathrm{diag}(e^{-i\theta/2},e^{i\theta/2})$ is a virtual gate adding essentially the phase shift to next gates \cite{zgates}.

A native two-qubit gate in IBM Quantum Eagle family is Echoed Cross Resonance ($ECR$)
\cite{ecr}. One can transpile \cor{$CNOT$ with $ECR$ by} adding single-qubit gates.
The ECR gate acts on the state $|ab\rangle$ as (Fig. \ref{ecr})
\ba
&ECR_\downarrow=((XI)-(YX))/\sqrt{2}=CR^- (XI) CR^+=\nonumber\\
&
\begin{pmatrix}
0&X_-\\
X_+&0\end{pmatrix}
=\begin{pmatrix}
0&0&1&i\\
0&0&i&1\\
1&-i&0&0\\
-i&1&0&0\end{pmatrix}/\sqrt{2},
\ea
in the basis $|00\rangle$, $|01\rangle$, $|10\rangle$, $|11\rangle$,
with Crossed Resonance gates
\be
CR^\pm=(ZX)_{\pm \pi/4}.
\ee
The gate is its inverse, i.e. $ECR_\downarrow ECR_\downarrow=(II)$.
It can be reversed, i.e., for $a\leftrightarrow b$, \cor{(see Fig. \ref{ecrr}):}
\be
ECR_\uparrow=((IX)-(XY))/\sqrt{2}=(HH) ECR_\downarrow(Y_+Y_-),
\ee
with Hadamard gate 
\be
H=(Z+X)/\sqrt{2}
=Z_+X_+Z_+=\begin{pmatrix}
1&1\\
1&-1\end{pmatrix}/\sqrt{2}
\ee
and $Z_\pm X_+Z_\mp=Y_\pm$, with $Y_+=HZ$ and $Y_-=ZH$.
The CNOT gate can be expressed by $ECR$ (Fig. \ref{cnot})
\ba
&CNOT_\downarrow=((II)+(ZI)+(IX)-(ZX))/2=\nonumber\\
&
\begin{pmatrix}
I&0\\
0&X\end{pmatrix}
=\begin{pmatrix}
1&0&0&0\\
0&1&0&0\\
0&0&0&1\\
0&0&1&0\end{pmatrix}=
(Z_+ I)ECR_\downarrow (XX_+),
\ea
while its reverse reads (Fig. \ref{cnotr})
\ba
&CNOT_\uparrow=((II)+(IZ)+(XI)-(XZ))/2
=\nonumber\\
&\begin{pmatrix}
1&0&0&0\\
0&0&0&1\\
0&0&1&0\\
0&1&0&0\end{pmatrix}=\nonumber\\
&
(HH)CNOT_\downarrow(HH)=\nonumber\\
&
(HH)ECR_\downarrow (X_+X_+)(Z_-H).
\ea

\begin{figure}
\begin{tikzpicture}[scale=1]
		\begin{yquantgroup}
			\registers{
			qubit {} q[2];
			}
			\circuit{
			box {$\uparrow$} (q[0,1]);
			}
			\equals
			\circuit{
			box {$Y_+$}  q[0];
			box {$Y_-$}   q[1];
			box {$\downarrow$}  (q[0,1]);
			box {$H$}  q[0];
			box {$H$}   q[1];
			}
		\end{yquantgroup}
\end{tikzpicture}

\caption{The $ECR_\uparrow$ gate expressed by $ECR_\downarrow$ }
\label{ecrr}
\end{figure}

\begin{figure}
\begin{tikzpicture}[scale=1]
		\begin{yquantgroup}
			\registers{
			qubit {} q[2];
			}
			\circuit{
			box {$X$}  q[0];
			box {$X_+$}   q[1];
			box {$\downarrow$}  (q[0,1]);
			box {$Z_+$}  q[0];
			}
			\equals
			\circuit{
			cnot q[1] | q[0];
			}
		\end{yquantgroup}
\end{tikzpicture}

\caption{The $CNOT_\downarrow$ gate expressed by $ECR_\downarrow$. }
\label{cnot}
\end{figure}
 
 \begin{figure}
\begin{tikzpicture}[scale=1]
		\begin{yquantgroup}
			\registers{
			qubit {} q[2];
			}
			\circuit{
			box {$Z_-$}  q[0];
			box {$H$}   q[1];
			box {$X_+$}  q[0];
			box {$X_+$}   q[1];
			box {$\downarrow$}  (q[0,1]);
			box {$H$}  q[0];
			box {$H$}  q[1];
			}
			\equals
			\circuit{
			box {$H$}  q[0];
			box {$H$}   q[1];
			cnot q[1] | q[0];
			box {$H$}  q[0];
			box {$H$}   q[1];
			}
			\equals
			\circuit{
			cnot q[0] | q[1];
			}
		\end{yquantgroup}
\end{tikzpicture}

\caption{The $CNOT_\uparrow$ gate expressed by $ECR_\downarrow$. }
\label{cnotr}
\end{figure}

In the newest, Heron family devices (\emph{ibm\_torino}, \emph{ibm\_fez}, \emph{ibm\_pittsburgh}), the native entangling gate is $CZ$. Its
action is very simple
\begin{equation}
CZ|ab\rangle=(-1)^{ab}|ab\rangle
\end{equation}
i.e. it reverses the sign of $|11\rangle$ component, or equivalently $CZ=(II+IZ+ZI-ZZ)/2$.
Both $CNOT$ gates can be just obtained by \cor{ applying $H$ gates to the second system, prior and after applying $CZ$ gate:}
\begin{equation}
CNOT_\downarrow=(IH)CZ(IH).
\end{equation}
\cor{We also depict it} in Fig.~\ref{cztran}. 

\begin{figure}
\begin{tikzpicture}[scale=1]
		\begin{yquantgroup}
			\registers{
			qubit {} q[2];
			}
			\circuit{
			box {$H$}  q[1];
			zz  (q[0,1]);
			box {$H$}  q[1];
			}
			\equals
			\circuit{
			cnot q[1] | q[0];
			}
		\end{yquantgroup}
\end{tikzpicture}

\caption{The $CNOT_\downarrow$ gate expressed by $CZ$ (the vertical link) and Hadamard \cor{gates} $H$.}
\label{cztran}
\end{figure}


\end{document}